\documentclass[twoside,twocolumn,9pt]{article}
\usepackage{extsizes}
\usepackage[super,sort&compress,comma]{natbib} 
\usepackage[version=3]{mhchem}
\usepackage[left=1.5cm, right=1.5cm, top=1.785cm, bottom=2.0cm]{geometry}
\usepackage{balance}
\usepackage{widetext}
\usepackage{times,mathptmx}
\usepackage{sectsty}
\usepackage{graphicx} 
\usepackage{lastpage}
\usepackage[format=plain,justification=raggedright,singlelinecheck=false,font={stretch=1.125,small,sf},labelfont=bf,labelsep=space]{caption}
\usepackage{float}
\usepackage{fancyhdr}
\usepackage{fnpos}
\usepackage[english]{babel}
\usepackage{array}
\usepackage{droidsans}
\usepackage{charter}
\usepackage[T1]{fontenc}
\usepackage[usenames,dvipsnames]{xcolor}
\usepackage{setspace}
\usepackage[compact]{titlesec}
\usepackage{amsmath}
\usepackage{amssymb}

\usepackage{epstopdf}

\definecolor{cream}{RGB}{222,217,201}

\begin{document}

\pagestyle{fancy}
\thispagestyle{plain}
\fancypagestyle{plain}{

\fancyhead[C]{\includegraphics[width=18.5cm]{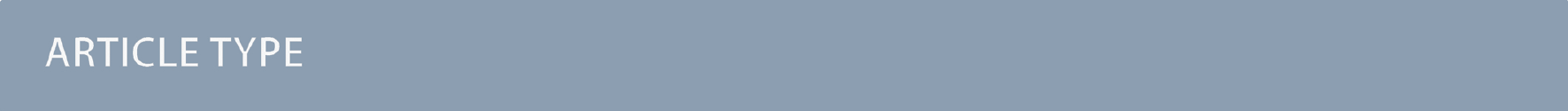}}
\fancyhead[L]{\hspace{0cm}\vspace{1.5cm}\includegraphics[height=30pt]{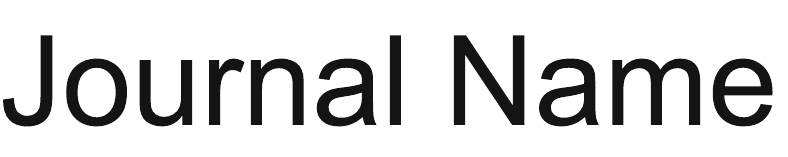}}
\fancyhead[R]{\hspace{0cm}\vspace{1.7cm}\includegraphics[height=55pt]{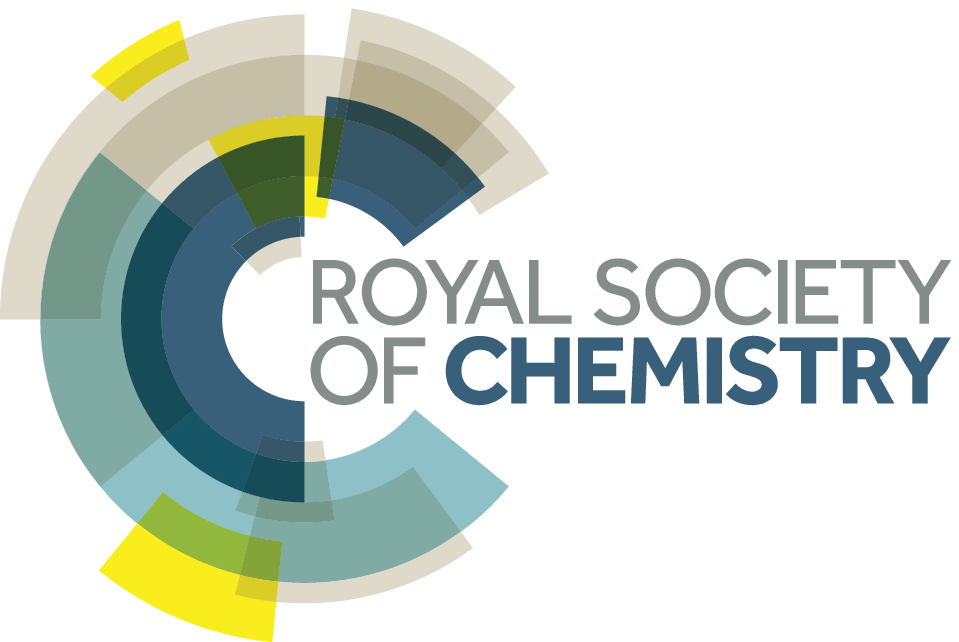}}
\renewcommand{\headrulewidth}{0pt}
}

\makeFNbottom
\makeatletter
\renewcommand\LARGE{\@setfontsize\LARGE{15pt}{17}}
\renewcommand\Large{\@setfontsize\Large{12pt}{14}}
\renewcommand\large{\@setfontsize\large{10pt}{12}}
\renewcommand\footnotesize{\@setfontsize\footnotesize{7pt}{10}}
\makeatother

\renewcommand{\thefootnote}{\fnsymbol{footnote}}
\renewcommand\footnoterule{\vspace*{1pt}%
\color{cream}\hrule width 3.5in height 0.4pt \color{black}\vspace*{5pt}} 
\setcounter{secnumdepth}{5}

\makeatletter 
\renewcommand\@biblabel[1]{#1}            
\renewcommand\@makefntext[1]%
{\noindent\makebox[0pt][r]{\@thefnmark\,}#1}
\makeatother 
\renewcommand{\figurename}{\small{Fig.}~}
\sectionfont{\sffamily\Large}
\subsectionfont{\normalsize}
\subsubsectionfont{\bf}
\setstretch{1.125} 
\setlength{\skip\footins}{0.8cm}
\setlength{\footnotesep}{0.25cm}
\setlength{\jot}{10pt}
\titlespacing*{\section}{0pt}{4pt}{4pt}
\titlespacing*{\subsection}{0pt}{15pt}{1pt}

\fancyfoot{}
\fancyfoot[LO,RE]{\vspace{-7.1pt}\includegraphics[height=9pt]{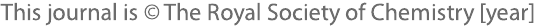}}
\fancyfoot[CO]{\vspace{-7.1pt}\hspace{13.2cm}\includegraphics{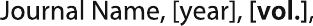}}
\fancyfoot[CE]{\vspace{-7.2pt}\hspace{-14.2cm}\includegraphics{head_foot/RF}}
\fancyfoot[RO]{\footnotesize{\sffamily{1--\pageref{LastPage} ~\textbar  \hspace{2pt}\thepage}}}
\fancyfoot[LE]{\footnotesize{\sffamily{\thepage~\textbar\hspace{3.45cm} 1--\pageref{LastPage}}}}
\fancyhead{}
\renewcommand{\headrulewidth}{0pt} 
\renewcommand{\footrulewidth}{0pt}
\setlength{\arrayrulewidth}{1pt}
\setlength{\columnsep}{6.5mm}
\setlength\bibsep{1pt}

\makeatletter 
\newlength{\figrulesep} 
\setlength{\figrulesep}{0.5\textfloatsep} 

\newcommand{\topfigrule}{\vspace*{-1pt}%
\noindent{\color{cream}\rule[-\figrulesep]{\columnwidth}{1.5pt}} }

\newcommand{\botfigrule}{\vspace*{-2pt}%
\noindent{\color{cream}\rule[\figrulesep]{\columnwidth}{1.5pt}} }

\newcommand{\dblfigrule}{\vspace*{-1pt}%
\noindent{\color{cream}\rule[-\figrulesep]{\textwidth}{1.5pt}} }

\makeatother

\twocolumn[
  \begin{@twocolumnfalse}
\vspace{3cm}
\sffamily
\begin{tabular}{m{4.5cm} p{13.5cm} }


\includegraphics{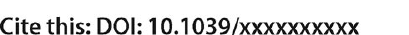} & \noindent\LARGE{\textbf{Mechanisms for collective inversion-symmetry breaking in dabconium perovskite ferroelectrics$^\dag$}} \\
\vspace{0.3cm} & \vspace{0.3cm} \\

 & \noindent\large{Dominic J. W. Allen,$^{\textsf{a}}$ Nicholas C. Bristowe,$^{\textsf{b}}$ Andrew L. Goodwin$^{\textsf{a},\ast}$  and Hamish H.-M. Yeung$^{\textsf{a,c}\ast}$} \\

\includegraphics{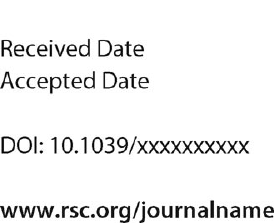} & \noindent\normalsize{Dabconium hybrid perovskites include a number of recently-discovered ferroelectric phases with large spontaneous polarisations. The origin of ferroelectric response has been rationalised in general terms in the context of hydrogen bonding, covalency, and strain coupling. Here we use a combination of simple theory, Monte Carlo simulations, and density functional theory calculations to assess the ability of these microscopic ingredients---together with the always-present through-space dipolar coupling---to account for the emergence of polarisation in these particular systems whilst not in other hybrid perovskites. Our key result is that the combination of A-site polarity, preferred orientation along $\langle111\rangle$ directions, and ferroelastic strain coupling drives precisely the ferroelectric transition observed experimentally. We rationalise the absence of polarisation in many hybrid perovskites, and arrive at a set of design rules for generating FE examples beyond the dabconium family alone.} \\

\end{tabular}

 \end{@twocolumnfalse} \vspace{0.6cm}
]
  

\renewcommand*\rmdefault{bch}\normalfont\upshape
\rmfamily
\section*{}
\vspace{-1cm}


\footnotetext{\textit{$^{\textsf{a}}$Department of Chemistry, University of Oxford, Inorganic Chemistry Laboratory, South Parks Road, Oxford OX1 3QR, U.K.; Tel: +44 1865 272137; E-mail: andrew.goodwin@chem.ox.ac.uk\\
$^{\textsf{b}}$School of Physical Sciences, University of Kent, Canterbury CT2 7NH, U.K.\\
$^{\textsf{c}}$School of Chemistry, University of Birmingham, Edgbaston, Birmingham B15 2TT, U.K.; Email: h.yeung@bham.ac.uk}}

\footnotetext{\dag~Electronic Supplementary Information (ESI) available. See DOI: 10.1039/b000000x/}




\section{Introduction}

It is a remarkable recent discovery that some dabconium hybrid perovskites are ferroelectric (FE) with spontaneous polarisations comparable to that of BaTiO$_3$.\cite{Ye_2018,Zhang_2017,Merz_1953} Remarkable, because polarisation is a measure of displaced charge density, and both the magnitude of localised charges and ABX$_3$ density are substantially reduced in hybrid perovskites relative to their longer-established inorganic cousins.\cite{Li_2017,Kieslich_2017} The possibility of combining the electrical performance of conventional ceramics with the mechanical flexibility and low-temperature processibility of hybrids is exciting because it offers many potential advantages in the development of \emph{e.g.}\ wearable electronics, flexible devices, and bionics.\cite{You_2017,Ye_2018,Li_2018,Guo_2019}

\begin{figure}
 \centering
\includegraphics{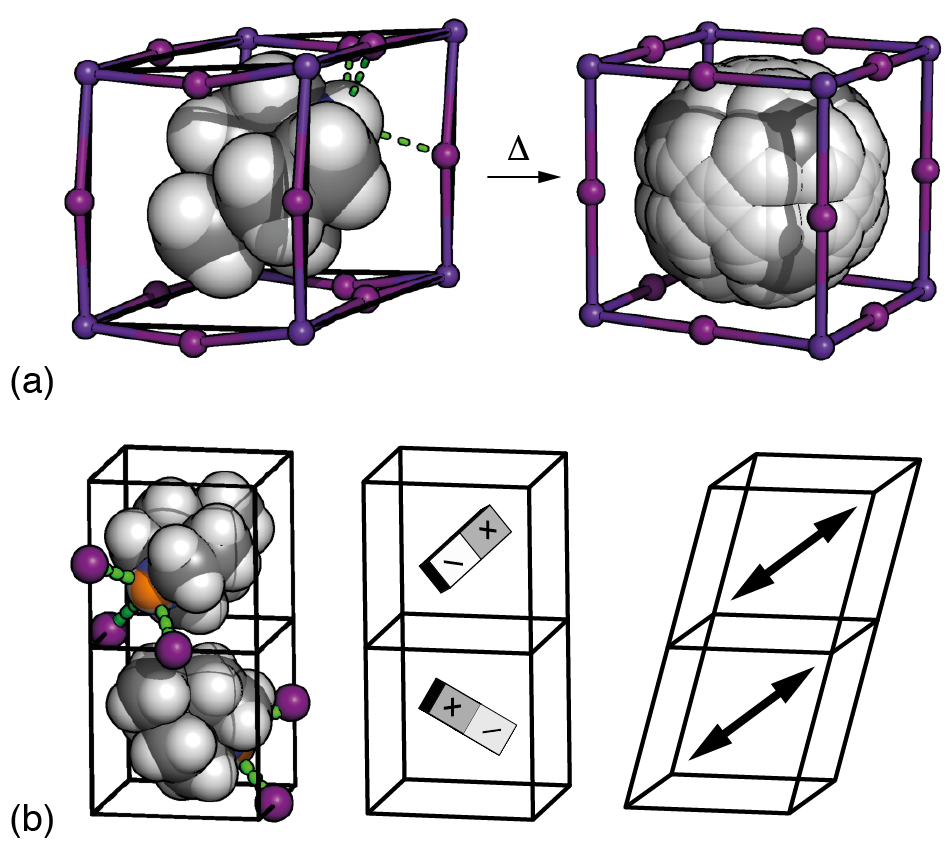}
 \caption{(a) The FE phase transition in [MDABCO]RbI$_3$ involves orientational ordering of MDABCO ions (space filling representation) on the A-site of the ABX$_3$ perovskite lattice. The low-temperature form has rhombohedral $R3$ symmetry, and the disordered high-temperature (paraelectric) form has $Pm\bar3m$ symmetry.\cite{symnote} (b) Candidate microscopic driving forces for FE order: (left--right) the tendency for each iodide anion (purple spheres) to hydrogen-bond with a single N-H proton (orange sphere), dipolar interactions, and strain coupling.}
 \label{fig1}
\end{figure}

The basic phenomenology of this family is well exemplified by [MDABCO]RbI$_3$ (MDABCO$^{2+}$ = {\it N}-methyl-{\it N}$^\prime$-diazabicyclo-[2.2.2]octonium) [Fig.~\ref{fig1}(a)].\cite{Zhang_2017} At high temperatures this system adopts the aristotypic (cubic) ABX$_3$ perovskite structure, with the high crystal symmetry reflecting orientational disorder of polar MDABCO A-site cations. On cooling below $\sim$400\,K an orientational ordering transition occurs such that each MDABCO cation now aligns along a common $\langle111\rangle$ direction, giving a structure characterised by the polar space-group $R3$. This phase is FE as its polarisation can be reversed under an applied field. The high-profile ``metal-free'' congener [MDABCO](NH$_4$)I$_3$ behaves analogously, but with NH$_4^+$ orientations amplifying the saturation polarisation in the FE phase.\cite{Ye_2018,Ehrenreich_2019}

Since relatively few amongst the diverse and extensive family of hybrid perovskites are polar,\cite{Li_2017,Gebhardt_2019,Ning_2019} it is natural to question what makes this particular family so special. Ultimately, of course, the goal is to develop design principles that allow the targeted synthesis and optimisation of hybrid ferroelectrics. On the simplest level, the use of polar A-site cations is considered important: indeed this was a key conclusion of Ref.~\citenum{Zhang_2017} on finding that replacing the $C_{3v}$-symmetric [MDABCO]$^{2+}$ cation by [DABCO]$^{2+}$ (= {\it N,N}$^\prime$-diazabicyclo-[2.2.2]octonium, $D_{3h}$ point symmetry), the corresponding perovskites do not have polar ground states. This principle has now matured into the so-called ``quasispherical theory'', whereby the local polarisation arising from asymmetric substitution of spherical cations is clearly linked empirically to FE response in perovskites and beyond:\cite{Zhang_2019,Yang_2019,Morita_2019,Wei_2020} \emph{e.g.}\ the DABCO cation is non-polar and roughly spherical, while MDABCO is polar and ovoid. But if A-site polarisation is a necessary condition it is not a sufficient one: many hybrid perovskites with polar A-site cations---from the famous photovoltaic [CH$_3$NH$_3$]PbI$_3$ to the antiferroelectric (AFE) multiferroics [NH$_2$(CH$_3$)$_2$]M(HCOO)$_3$---are not themselves polar.\cite{Weller_2015,Weller_2015b,Jain_2008}


So the question of what couples dipole orientations in dabconium perovskites to generate bulk polarisation is an important and open problem.\cite{Shahrokhi_2020} The elongated shape of MDABCO itself is understood to favour alignment along $\langle111\rangle$, with this orientation also allowing hydrogen bonding between the R$_3$N--H proton and three nearby halide ions held in a \emph{fac} arrangement around a common B-site cation.\cite{Zhang_2017} The electronic implications of this hydrogen bonding---in essence a requirement that each halide interact strongly with only a single MDABCO cation---and the steric implications of coupling between distortions of the [BX$_3$]$^{2-}$ lattice have been proposed in general terms as coupling mechanisms [Fig.~\ref{fig1}(b)].\cite{Zhang_2017,Ye_2018} The logical difficulty one faces is that similar considerations are at play in most hybrid perovskites, including the many non-polar examples. 






Here we seek to understand whether the different microscopic ingredients of hydrogen-bonding rules, distortion strain coupling, and the always-present through-space dipolar interactions---either on their own or in combination---can account for the observed FE transition observed experimentally in dabconium perovskites. We use a coarse-graining approach supplemented by straightforward calculations and Monte Carlo (MC) simulations to identify ground states for various combinations of these ingredients. Our key result is that the pairing of strain coupling and dipolar interactions stabilises the $R3$ polar ground state when the two interactions are not too dissimilar in strength. Using density functional theory (DFT) calculations on a range of suitably-chosen [MDABCO]RbI$_3$ polymorphs, we demonstrate that our simple theoretical model provides a sensible coarse graining of the energetics of this particular system, and hence provides a first-order explanation for its FE behaviour. Finally, we show that this crucial dipolar--strain combination drives collective polarisation only when there is a driving force for alignment of the A-site dipole along $\langle111\rangle$; the corresponding ground states for $\langle100\rangle$ and $\langle110\rangle$ are non-polar. In this way we rationalise the absence of polarisation in many hybrid perovskites, and arrive at a detailed set of design rules for generating FE examples beyond the dabconium family alone.


\section{Results}

\begin{figure}[b]
 \centering
 \includegraphics{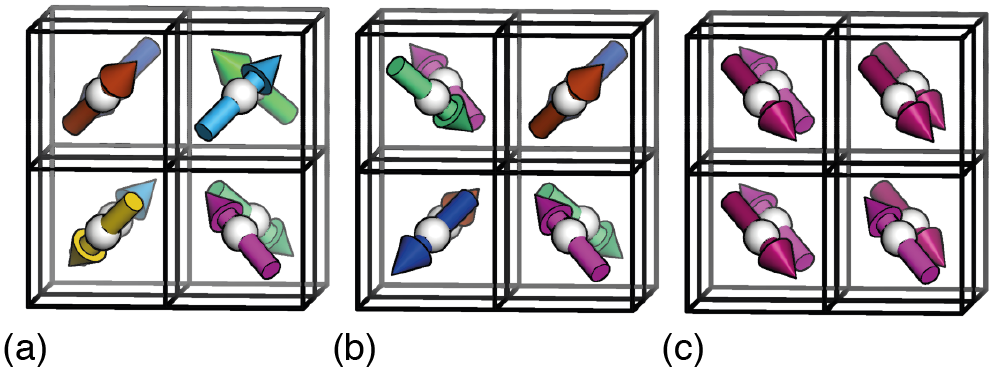}
 \caption{Representative ground states for each of the interactions depicted in Fig.~\ref{fig1}(b). Arrows denote the orientations of MDABCO $C_{3v}$ axes (\emph{i.e.}\ N--H bond vectors) and are coloured accordingly. (a) Hydrogen-bonding interactions give a correlated-disordered ground state with $Pm\bar3m$ symmetry. (b) Dipolar interactions drive an ordered state with $I23$ symmetry. (c) Strain coupling gives a partially-ordered (nematic) state, the average structure of which has $R\bar3$ symmetry. None of the three states is polar.}
 \label{fig2}
\end{figure}

Our starting point is to consider in turn the implications of each of these three interaction types by themselves. Taking first the case of hydrogen-bonding, we recall that each MDABCO$^{2+}$ ion forms its strongest hydrogen bonds with a triplet of halide ions that share a triangular face of the AX$_{12}$ cuboctahedron (see SI for further discussion).\cite{Zhang_2017,Ye_2018} Hydrogen bonding reduces the effective charge on each of these three anions, such that they are now less available for additional bonding to the A-site cations of neighbouring perovskite cages. Hence the orientation of one MDABCO constrains those of its neighbours, generating a `hydrogen-bonding rule' that couples MDABCO orientations and at face value may imply long-range FE order. We identify the implications for crystal symmetry by carrying out MC simulations driven by the coarse-grained energy
\begin{equation}\label{hamil1}
E_{\rm HB}=H\sum_{j\in\{\rm X\}}(n_j-1)^2.
\end{equation}
Here the sum is over all X-site anions in the MC configuration (a supercell of the aristotypic ABX$_3$ cell), $n_j$ is the number of MDABCO cations strongly hydrogen-bonded to the $j^{\rm th}$ anion, and $H>0$ is the energy penalty for double-bonding. The ground state of Eq.~\eqref{hamil1} is the set of configurations for which each anion is hydrogen-bonded to a single MDABCO cation, and the form of this equation can be rationalised as the leading term in the expansion of the crystal energy around this minimum. Our MC simulations show the ground state to be disordered---but strongly correlated---such that its average crystal symmetry remains $Pm\bar3m$ [Fig.~\ref{fig2}(a)]. There is a strong conceptual parallel to the disordered `chain' structure of paraelectric BaTiO$_3$, where Ti--O covalency drives a strongly-correlated nonpolar state.\cite{Comes_1970,Senn_2016}

In a similar manner, MC simulations also allow us to test the symmetry-breaking implications of dipole--dipole interactions. The corresponding configurational energy is now
\begin{equation}\label{hamil2}
E_{\rm dip}=D\sum_{i\ne j}\frac{\mathbf S_i\cdot\mathbf S_j-3(\mathbf S_i\cdot\hat{\mathbf r}_{ij})(\mathbf S_j\cdot\hat{\mathbf r}_{ij})}{(r_{ij}/a)^3},
\end{equation}
where $D$ is the dipolar interaction strength and the sum is over all distinct A-site cations $i,j$ with MDABCO orientations $\mathbf S_i,\mathbf S_j\in\langle111\rangle$. The vector $\hat{\mathbf r}_{ij}$ is the normalised vector between the two cations, $r_{ij}$ their absolute separation, and $a$ the unit-cell length. Our MC implementation\cite{Paddison_2015} takes into account the need for Ewald summation as dipolar interactions are long-range. We find the ground state of Eq.~\eqref{hamil2} to be the fully-ordered but non-polar arrangement with $I23$ crystal symmetry shown in Fig.~\ref{fig2}(b). This particular arrangement, the generalised form of which is known from theory,\cite{Belobrov_1983} is related to that found in the low-temperature phases of dipolar solids such as CO$_{({\rm s})}$.\cite{Lipscomb_1974}

The third and final interaction type is that of strain coupling. Formally, strain is a rank-two tensor, but because the ovoid shape of the MDABCO cation strains the corresponding perovskite cell along the same $\langle111\rangle$ axis along which the MDABCO is oriented, we can simplify strain coupling in terms of the biquadratic interaction\cite{Coates_2019}
\begin{equation}\label{hamil3}
E_{\rm strain}=-J\sum_{i,j}(\mathbf S_i\cdot\mathbf S_j)^2.
\end{equation}
This sum is over neighbouring A-sites $i,j$, with orientations $\mathbf S_i,\mathbf S_j\in\langle111\rangle$ as above. The magnitude of $J$ determines the strength of strain coupling, and if $J>0$ (as it likely is for the dabconium perovskites; see further discussion below) this coupling is ferroelastic. The ground state of Eq.~\eqref{hamil3} is obvious by inspection: it is the set of configurations with all A-site cations oriented parallel or antiparallel to the same $\langle111\rangle$ axis. This state is disordered with $R\bar3$ symmetry [Fig.~\ref{fig2}(c)] and so again is non-polar. In the context of magnetic and liquid-crystal statistical mechanics, this is a `nematic' phase with quadrupolar (\emph{cf} strain)---but not dipolar---order.


Consequently, we can conclude that none of the three microscopic interactions proposed in the literature can by themselves account for inversion-symmetry breaking in the dabconium perovskites. Hence, we consider the different combinations of the interactions in turn, in order to examine whether applied together they explain the polar $R3$ ground state observed experimentally.



\begin{figure}
 \centering
 \includegraphics{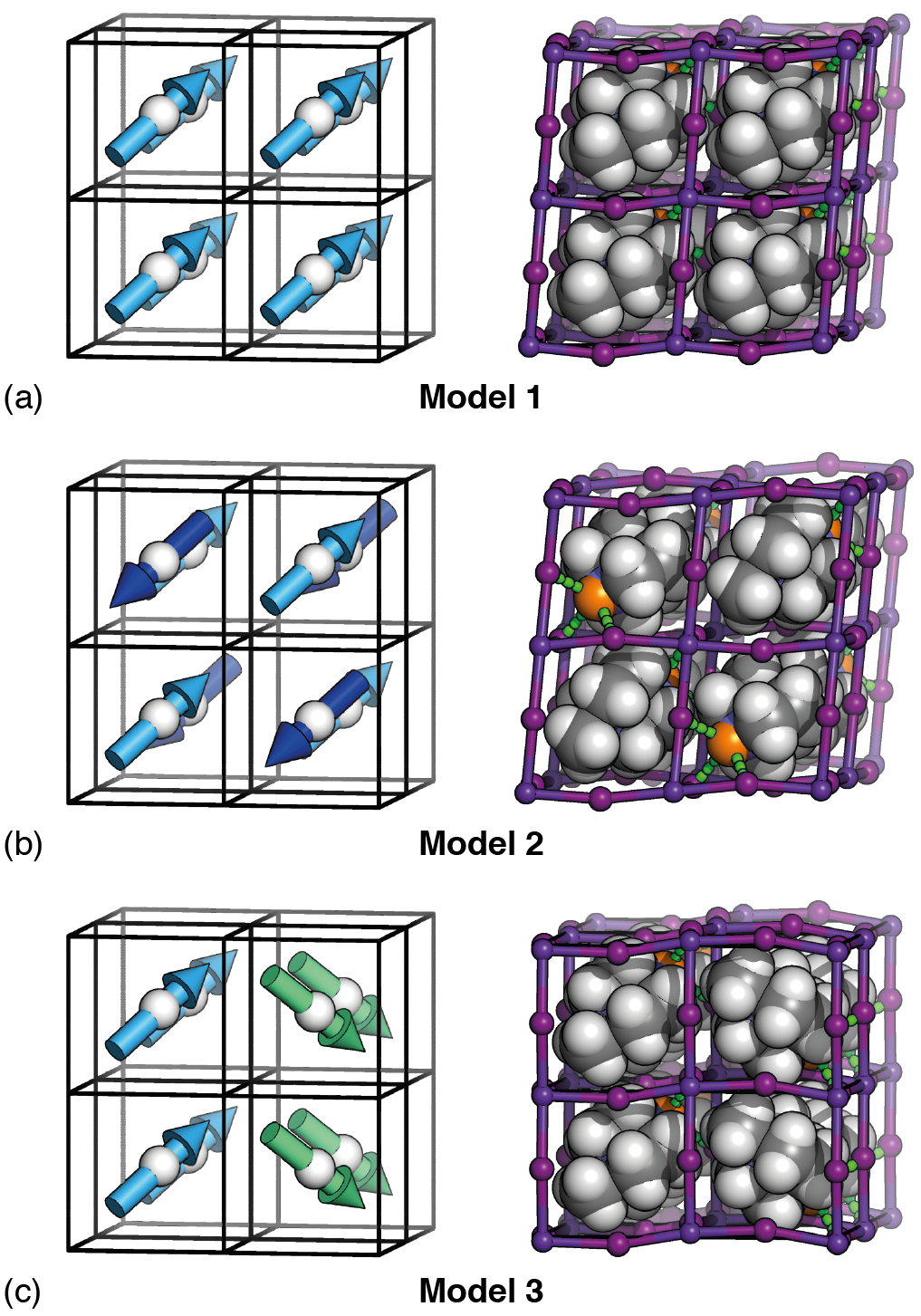}
 \caption{Three model systems used to consider the importance of hydrogen-bonding rules in stabilising the polar $R3$ phase. (a) {\bf Model 1:} The FE $R3$ phase itself. (b) {\bf Model 2:}  An (R-point) AFE variant with $R\bar3$ symmetry; note that the dipole orientation alternates from cell to cell. (c) {\bf Model 3:} A ferrielectric arrangement in which successive planes switch one component of their polarisation. In each case the underlying dipole arrangements are shown on the left, and the corresponding DFT-relaxed structures are shown on the right. All models satisfy the hydrogen-bonding rule that each iodide anion is strongly bonded to one and only one MDABCO cation.}
 \label{fig3}
\end{figure}

We next show that hydrogen-bonding rules, applied in conjunction with dipole--dipole interactions or strain coupling, also fail to produce a macroscopic polarisation. To do this, we compare
 three very simple (ordered) models of MDABCO orientations that all strictly obey the hydrogen-bonding rules discussed above (\emph{i.e.} $E_{\rm{HB}}=0$) [Fig.~\ref{fig3}]. Model 1 is the $R3$ structure itself; Model 2 is an AFE variant of this structure with $R\bar3$ symmetry where the MDABCO orientation alternates in a checkerboard fashion from cell to cell (R-type AFE order); and Model 3 is a ``ferri''-electric $2\times1\times1$ supercell of the $R3$ structure with monoclinic $Pc$ symmetry in which neighbouring planes of A-site cations point alternately along $[111]$ and $[11\bar1]$ axes.

Since Models 1 and 2 involve alignment of A-site cations parallel and antiparallel to a single common axis, the corresponding $E_{\rm strain}$ terms are identical, irrespective of the value of $J$. Hence both are ground states of any interaction model based on hydrogen bonding and strain coupling together---and there is nothing to favour the observed $R3$ state over the $R\bar3$ AFE alternative. By a similar argument, the dipolar interaction energy of Model 3 ($E_{\rm dip}=-2.204D$; determined by numerical evaluation of Eq.~\eqref{hamil2} using Ewald summation) is always more favourable than that of Model 1 ($E_{\rm dip}=-2.094D$) irrespective of the value of $D$. Thus, interaction models based on hydrogen bonding and dipole--dipole interactions are also unable to stabilise the $R3$ state.

\begin{figure}
 \centering
\includegraphics{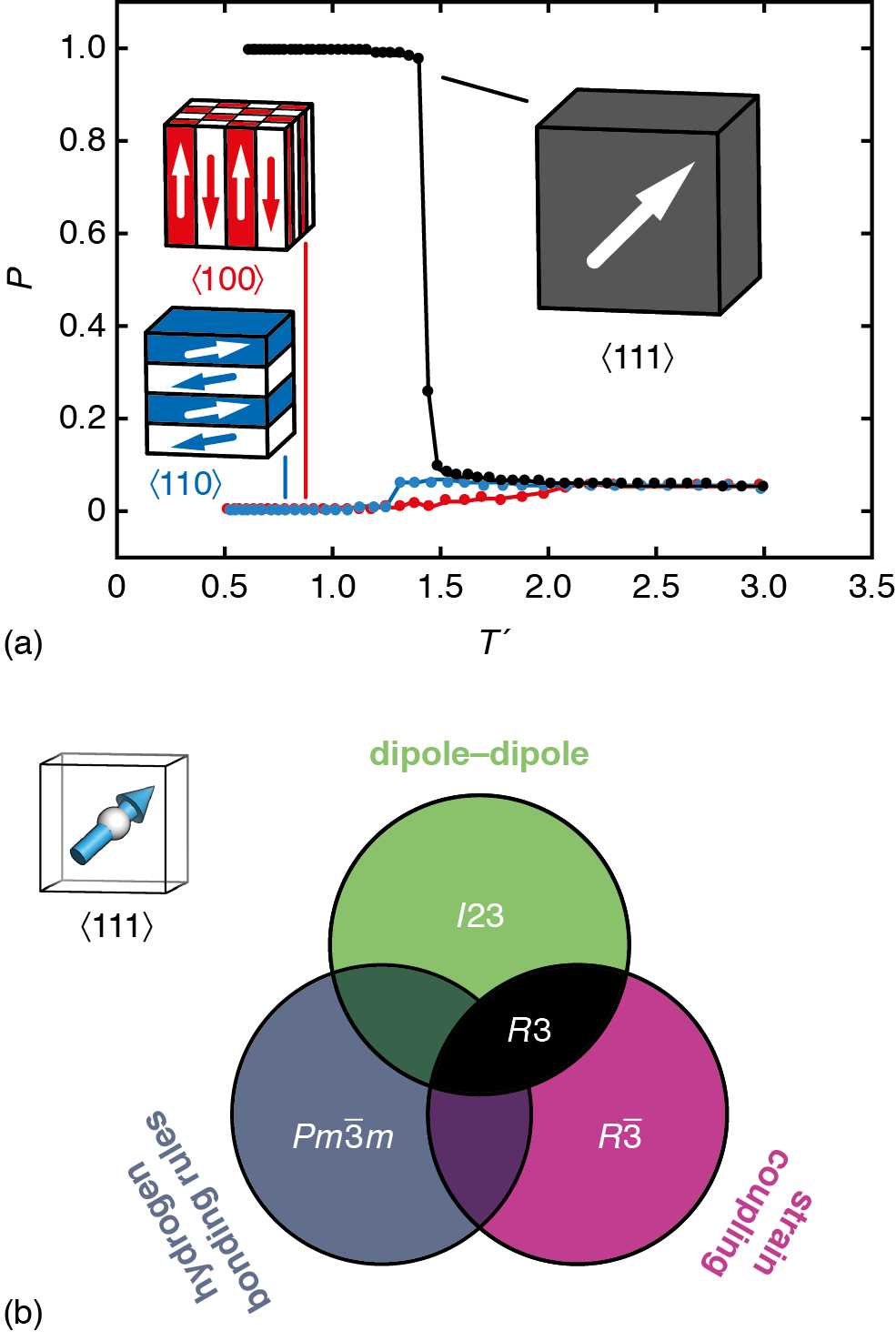}
 \caption{Emergence of spontaneous polarisation in the ground state of Eq.~\eqref{dualhamil} (a) Temperature dependence of the bulk polarisation as determined using Monte Carlo simulation. The three traces correspond to the simulations in which local orientations are confined to $\langle111\rangle$ (black), $\langle110\rangle$ (blue), and $\langle100\rangle$ axes. The corresponding ground states are represented schematically; only that of the $\langle111\rangle$ system is polar. Error bars are smaller than the symbols. (b) Summary of the symmetry-breaking implications of the three microscopic orientation-coupling mechanisms discussed in the text. For perovskites with polar B-site cations aligned preferentially along $\langle111\rangle$ axes, the combination of strain-coupling and dipole--dipole interactions (black region) is sufficient to stabilise the polar $R3$ ground state observed experimentally in FE dabconium perovskites.}
 \label{fig4}
\end{figure}

By contrast, the combination of dipole--dipole interactions and strain-coupling \emph{does} give a polar ground state with $R3$ symmetry. We obtain this result using MC simulations driven by the combined interaction energy
\begin{equation}\label{dualhamil}
E_{\rm MC}=E_{\rm dip}+E_{\rm strain},
\end{equation}
setting $J\simeq D$. Fig.~\ref{fig4}(a) illustrates our finding that this model exhibits a phase transition at $T_{\rm c}\sim1.5D(,J)$ by tracking the average relative polarisation $P=|\langle\mathbf S\rangle|$ as a function of reduced temperature $T^\prime=T/D$; the low-temperature phase is precisely the $R3$ state. This polar phase is the ground state for all finite $D$ until $D\gtrsim4.6J$, beyond which the $I23$ phase is the more stable. Moreover, since the $R3$ state obeys our hydrogen-bonding rules, it is stable regardless of the importance or otherwise of an additional $E_{\rm HB}$ term. We summarise in Fig.~\ref{fig4}(b) this key result of our analysis. The emergence of polarisation \emph{via} the interplay of two competing interactions---neither of which can by itself break inversion symmetry---has strong conceptual parallels to the phenomenology of hybrid-improper ferroelectrics.\cite{Bousquet_2008,Benedek_2011,Wolpert_2019}

\begin{figure}
 \centering
 \includegraphics{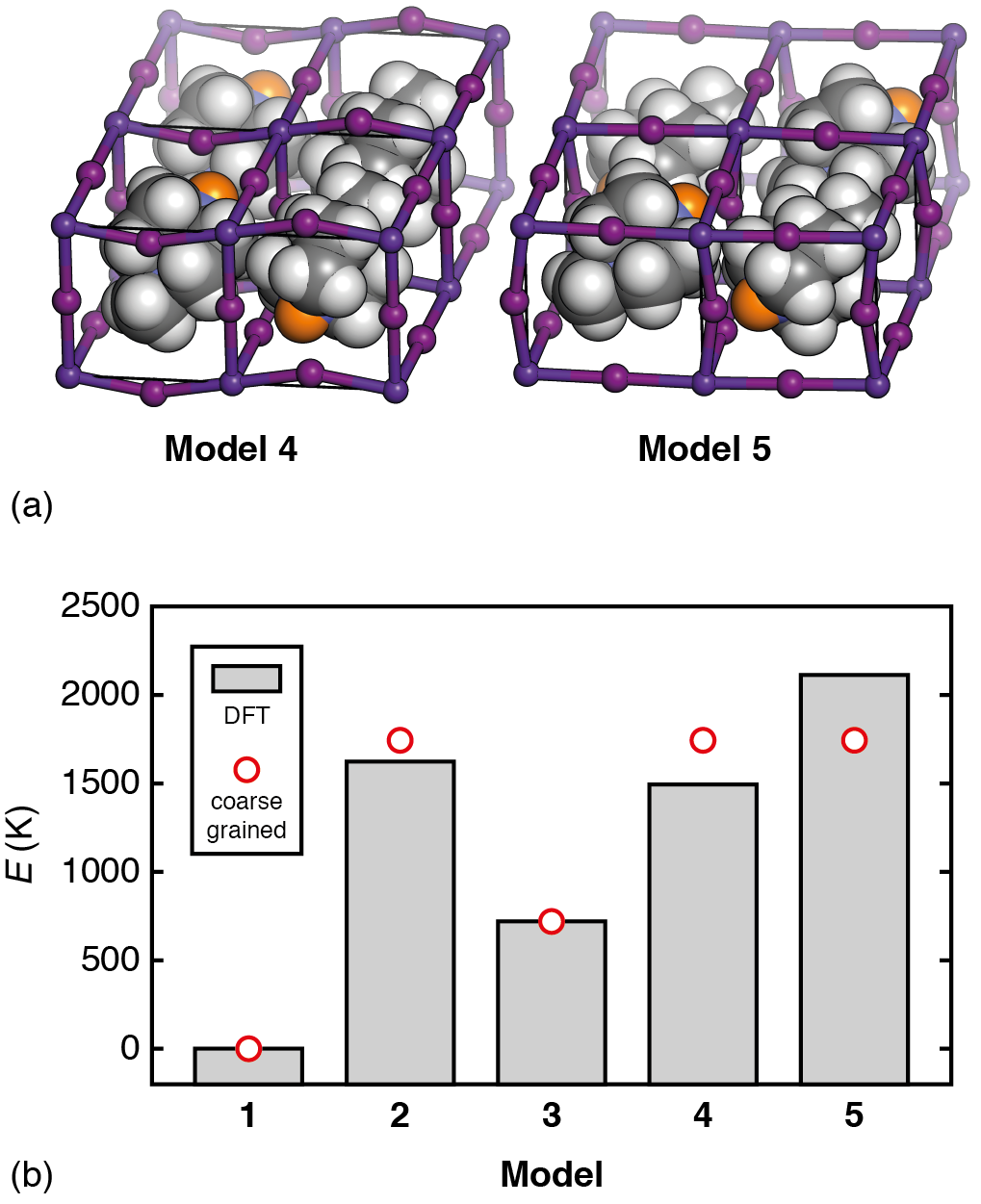}
 \caption{(a) The additional AFE models of [MDABCO]RbI$_3$ used to determine the relative importance of dipolar and strain interactions as described in the text. {\bf Model 4:} An X-point variant of the $R3$ FE ground state with triclinic $P\bar1$ space-group symmetry. {\bf Model 5:} The corresponding M-point AFE structure, which also has $P\bar1$ symmetry, albeit with a different unit cell. In both models, MDABCO orientations are parallel or antiparallel to the $[111]$ axis of the cubic aristotype. The structures shown are those obtained following DFT relaxation; we use the same representation as in Fig.~\ref{fig3}. (b) Comparison between \emph{ab initio} and coarse-grained energies for each of the models 1--5 of [MDABCO]RbI$_3$, given relative to the global energy minimum (Model 1). The reasonable match obtained suggests that the energetics of this system are well captured by the simple coarse-grained model, with similar energy scales for both dipolar and strain interactions.}
 \label{fig5}
\end{figure}

But how relevant is this coarse-grained model to a physical system such as [MDABCO]RbI$_3$? To answer this question we use DFT calculations to compare the energies of a range of [MDABCO]RbI$_3$ polymorphs with different MDABCO orientational ordering patterns. We consider five such configurations in total: Models 1--3 as described above, and the two simplest additional AFE orderings (formally, corresponding to X-point and M-point AFE order) [Fig.~\ref{fig5}(a)]. On the one hand, the coarse-grained energy of each model can be determined analytically in terms of the two parameters $J$ and $D$ (see SI). And, on the other hand, we can calculate the (real) energies using DFT. Equating the two we find good agreement for the values $D=833$\,K and $J=912$\,K; \emph{i.e.} $D\simeq J$ as above [see comparison of coarse-grained and DFT energies in Fig.~\ref{fig5}(b)]. Hence---to a first approximation---the energetics of this system are captured by the coarse graining implied by Eq.~\eqref{dualhamil}, which finally explains the driving force for FE order.

The variation in DFT energies amongst Models 2, 4, and 5---all of which have the same relative coarse-grained energy $E=+2.094D$---likely reflects the importance of deviations away from $\langle111\rangle $ in A-site orientations. The DFT-relaxed structure of model 4 in particular shows relatively large deviations of this type, which reduce the dipolar energy cost of its AFE ordering pattern and contribute to its stability relative to the other AFE models. Importantly, if we re-run our MC simulations allowing the vectors $\mathbf S_i$ to adopt any orientation---yet retaining some bias towards the $\langle111\rangle$ axes\cite{sianote}---the same FE transition is observed, albeit at lower critical temperature.\cite{tempnote} So strict adherence to $\langle111\rangle$-type orientational anisotropy is not required for inversion-symmetry breaking.



But some anisotropy of this type \emph{is} important. As a final calculation, we carried out a set of MC simulations using Eq.~\eqref{dualhamil} but with A-site orientations confined to either $\langle100\rangle$ or $\langle110\rangle$ orientations. While both systems undergo phase transitions on cooling, the ground states of the two models are AFE rather than FE [Fig.~\ref{fig4}(a)].\cite{Luttinger_1946,Schildknecht_2019} So the fact that MDABCO orients along the body diagonal of the perovskite cage---a function of its shape and hydrogen-bonding characteristics---is a crucial ingredient in its own right.


\section{Discussion and concluding remarks}

We might now claim to understand the origin of the FE phase in systems such as [MDABCO]RbI$_3$. The dipole moment of MDABCO$^{2+}$ is certainly important, but equally important is that this moment is aligned along a $\langle111\rangle$ body-diagonal of the perovskite cage. The displacement of MDABCO$^{2+}$ ions along this same axis amplifies the corresponding effective dipole moment.\cite{Wang_2019} The through-space interactions between neighbouring dipoles are always present, and turn out to be a crucial ingredient for collective polarisation. The final necessary component is strain coupling---the tendency for the strains in neighbouring cells to coalign. While hydrogen-bonding or covalency effects may help stabilise the FE state, they cannot explain its occurrence.


From a design perspective, there are two aspects one might wish to control---(i) A-site dipoles with a tendency to orient preferentially along $\langle111\rangle$, and (ii) ferroelastic strain coupling. The former is likely the easier (or at least more obvious): use ovoid A-site cations with three-fold rotational symmetry, small enough to stabilise the perovskite structure\cite{Mitzi_2001,Kieslich_2014,Burger_2018} but large enough to strain the perovskite cage along $\langle111\rangle$. The balance between ferroelastic and antiferroelastic strain coupling is less obviously navigated,\cite{Evans_2016} although we expect that for strains along $\langle111\rangle$ the ferroelastic case may be the more common. Antiferroelastic strain ($J<0$ in Eq.~\eqref{hamil3}) stabilises the same $I23$ ground state observed for dipolar interactions alone. An important distinction between this structure and the polar $R3$ structure is the extent of conformational degree of freedom of the BX$_6$ octahedra. In the ferroelastic $R3$ case, symmetry allows rotations of the BX$_6$ octahedra around the three-fold axis and distortions of their internal X--B--X angles---both of which help to accommodate the MDABCO$^{2+}$ ion and optimise hydrogen-bonding (as observed experimentally).\cite{Zhang_2017} By contrast, the B-site point symmetry of the antiferroelastic $I23$ structure forbids rotations and bond-angle distortions, allowing only modulation of the lengths of B--X bonds, which are much more expensive to distort.



With these design principles in mind, it is straightforward to rationalise why many well-known hybrid perovskites with polar A-site cations are nevertheless themselves apolar: the A-site dipoles are not aligned along $\langle111\rangle$ and/or the strain coupling is antiferroelastic. But the rules we develop are by no means exhaustive, since (obviously) other mechanisms can and do also drive FE order (see Refs.~\citenum{Pan_2017,Ferrandin_2019} and the ODABCO[NH$_4$]Cl$_3$ example in Ref.~\citenum{Ye_2018} for relevant counterexamples). So the key remaining challenge is to develop a more general understanding of which combinations of local symmetry breaking and various short- and long-range interactions can break inversion symmetry in order/disorder FE materials. 

\section*{Conflicts of interest}
There are no conflicts to declare.

\section*{Acknowledgements}
A.L.G. gratefully acknowledge the E.R.C. for funding (Advanced Grant 788144). H.H.-M.Y. thanks the University of Birmingham for Startup funds.




\renewcommand\refname{Notes and references}

\bibliography{mh_2020_abx3polar} 
\bibliographystyle{rsc} 

\end{document}